\documentclass[useAMS,usenatbib]{mn2e}
\usepackage{subfigure}
\usepackage{amsmath}
\usepackage{amssymb}
\usepackage{bm}
\usepackage{dcolumn}
\usepackage{graphicx}
\usepackage{color}
\usepackage{ulem}
\usepackage{dblfloatfix}

\newcommand       \apj          {ApJ}
\newcommand       \apjl         {ApJL}

\newcommand       \nat          {Nature}
\newcommand       \mnras        {MNRAS}

\makeatletter
\def\tbcaption{\def\@captype{table}\caption}
\def\figcaption{\def\@captype{figure}\caption}
\makeatother

\title[Beta-decay heating rate of macronova]
  {Analytic heating rate of neutron star merger ejecta derived from Fermi's theory of beta decay}
\author[K. Hotokezaka, R. Sari, and T. Piran]
  {Kenta Hotokezaka$^{1,2}$\thanks{E-mail: khotokezaka@simonsfoundation.org}, Re'em Sari$^1$, and Tsvi Piran$^1$
\\
  $^1$Racah Institute of Physics,~The Hebrew~University of Jerusalem,~Jerusalem,~91904,~Israel\\
  $^2$Center for Computational Astrophysics, 162 5th Ave, New York, NY, 10010, USA
  }
\date{10 January 2017}

\pagerange{\pageref{firstpage}--\pageref{lastpage}} \pubyear{2015}

\def\LaTeX{L\kern-.36em\raise.3ex\hbox{a}\kern-.15em
    T\kern-.1667em\lower.7ex\hbox{E}\kern-.125emX}

\begin{document}

\label{firstpage}

\maketitle

\begin{abstract}
Macronovae (kilonovae) that arise in binary neutron star mergers are powered by radioactive beta decay of 
hundreds of $r$-process nuclides. 
We derive, using Fermi's theory of beta decay, an analytic estimate of  the nuclear heating rate. We 
show that the heating rate evolves as a power law ranging between $t^{-6/5}$ to $t^{-4/3}$. 
The overall magnitude of the heating rate is determined by the mean values of nuclear 
quantities, e.g., the nuclear matrix elements of beta decay. These values are specified 
by using nuclear experimental data. We discuss the role of higher order beta transitions
and the robustness of the power law. 
The robust and simple form of the heating rate  suggests that 
observations of  the late-time bolometric light curve $\propto t^{-\frac{4}{3}}$
would be a  direct evidence of a $r$-process driven macronova.
Such  observations could also enable us to  estimate
the total amount of $r$-process nuclei produced in the merger.
\end{abstract}


\section{Introduction}
\label{sec:Introduction}

 \cite{li1998ApJ} proposed that neutron star mergers will be accompanied by 
macronovae (kilonovae), which are optical--infrared transients powered by radioactive decay 
of the merger's debris.
These macronovae are  among  the most promising electromagnetic 
counterparts to gravitational-wave merger events~(e.g. \citealt{metzger2012ApJ,nissanke2013ApJ}).
Recently, macronova candidates have been discovered in the afterglows of 
several short gamma-ray bursts~\citep{tanvir2013Nature, berger2013ApJ,yang2015NatCo,jin2016}.
\cite{jin2016} re-analyzed the afterglow light curves of
historical nearby short gamma-ray bursts and suggested 
that macronovae are ubiquitous in short GRBs' afterglows. 

The radioactive heat generated by $r$-process nuclei play an essential role in powering
macronovae.  Due to strong adiabatic cooling the ejecta's initial internal energy is
practically negligible at the time that the ejecta become optically thin, i.e. when $\tau \approx c/v$, where $\tau$ is the optical 
depth and $v$ is the velocity of the ejecta. As we discuss later, with
 typical parameters the  peak emission time 
is around a few days and hence this is the important timescale to focus on. 
Detailed computations using nuclear database and numerical 
simulations  have been  widely used to obtain the radioactive heating rates~\citep{freiburghaus1999ApJ,metzger2010MNRAS,goriely2011ApJ,roberts2011ApJ,
korobkin2012MNRAS, 
wanajo2014ApJ,lippuner2015ApJ,hotokezaka2016MNRAS,barnes2016}.

Another approach to calculate  the heating rate is to consider the 
$r$-process material as a statistical assembly of radioactive nuclei.
Incorporating Fermi's theory of beta decay with 
such an approach provides a clear physical understanding of the nuclear 
heating rate~(see, e.g., \citealt{Way1948} for a discussion on  the energy generation by fission products).
We follow this approach and use the Fermi theory to estimate the heating rate in neutron star mergers' ejecta. 
\citet[see also \citealt{metzger2010MNRAS}]{colgate1966ApJ} considered this approach to estimate the radioactive luminosity of supernovae.
But they used only the relativistic regime of Fermi's theory, which  is not relevant 
on the macronova peak timescale as we discuss later. Furthermore, they 
assumed that each element decays to a stable one rather than following
a decay chain, which we consider in this paper.



In this paper, we analytically derive the nuclear heating rate of macronovae 
based on Fermi's theory. In \S~\ref{sec:summary}, we begin with a brief summary of  the basic concepts
and the outcome of this work.
In \S~\ref{sec:fermi}, we introduce the key ingredients 
of the theory needed to calculate the heating rate. 
We derive the heating rate of the beta decay chains of allowed transitions in \S~\ref{sec:heat}.
We discuss, in \S~\ref{sec:discussion},  the role of forbidden transitions and other effects that
we  ignore and we estimate their possible effect. 
We summarize the results and discuss the implication to macronova studies in \S~\ref{sec:conclusion}.

\section{A short summary of beta-decay heating rate} \label{sec:summary}
Radioactive nuclei  that are far from the stability valley are produced in $r$-process nucleosynthesis.  
These nuclei undergo beta decay without changing their mass number. A series of
beta decays in each mass number  is considered as  a decay chain. Because
the mean lives of radioactive nuclides typically become longer when approaching 
the stability valley,  the nuclei in a decay chain at given time $t$
stay at some specific nuclide with a mean life $\tau \sim t$. 
This means that the number of decaying nuclei in a logarithmic
time interval is constant, i.e, the decay rate is $\sim N/t$, where
$N$ is the total  number of  nuclei in the chain.  Then the beta decay 
heating rate per nucleus is given by $\dot{q}\sim E(t)/t$, where $E(t)$ is 
the disintegration energy of the beta decay as a function of the mean life. 
As we will see later,  two important concepts in beta decay theory enable us to determine $E(t)$.
First, there are four physical constants in the problem, the Fermi's constant $G_{F}$,
the electron mass $m_e$, the speed of light $c$, and the Planck constant $\hbar$. 
The fundamental timescale of beta decay $t_{F} \approx 9\cdot 10^{3}$\,s can be obtained from these physical constants.
Second, there is a well known relation between the disintegration energy and
mean life as $\tau \propto E^{-5}$. Therefore, the heating rate per nucleus can be roughly estimated as
\begin{eqnarray}
\dot{q}(t)\sim \frac{m_e c^2}{t_F}\left(\frac{t}{t_{F}}\right)^{-1.2}.
\end{eqnarray}
This gives a correct order of magnitude and a reasonable estimate of the 
time dependence of the beta decay heating rate of $r$-process material.
In the following we refine these ideas.

\section{Basics of Fermi's theory  of beta decay} \label{sec:fermi}
The nature of beta decay was successfully described 
by \cite{Fermi1934}. Here we briefly describe key ingredients of Fermi's theory needed for obtaining the 
macronova heating rate. 
In a beta decay  one of the neutrons in a nucleus  disintegrated to a proton,
an electron, and an anti-neutrino.  Using  Fermi's Golden rule,
the beta-disintegration probability of a beta-unstable 
nucleus per unit time in a unit momentum interval of the electron is written as
\begin{eqnarray}
\frac{dw}{dp_{e}}dp_e = \frac{2\pi}{\hbar} \left| H_{fi}^{\prime}\right|^{2}\rho(E_{e}), \label{fermi}
\end{eqnarray}
where $p_{e}$ and $E_e$ are momentum and kinetic energy of the electron,
$H_{fi}^{\prime}$ is the matrix element of the interaction Hamiltonian responsible for 
the beta disintegration,
and $\rho(E_{e})$ is the number density of final states of the light particles 
between $E_e$ and $E_e+dE_e$. The number of final states
is assumed to be proportional to the volume of the accessible phase space of the light particles: 
\begin{eqnarray}
\rho(E_{e})dE_{e} & = &\frac{(4\pi)^{2}V^{2} }{(2\pi \hbar)^{6}} p_{e}^{2}dp_e p^{2}_{\nu} dp_{\nu} \nonumber \\
& \approx & \frac{(4\pi)^{2}V^{2}}{(2\pi \hbar)^{6}c^3}  \label{dos}
p_{e}^{2} dp_e (E_{0}-E_{e})^{2} dE_{e},
\end{eqnarray}
where $E_{0}$ is the total disintegration energy, $p_\nu$ is momentum of the neutrino,
and energy conservation  $E_0 \approx E_{e} + E_{\nu}$ has been  used.
We use the fact that  the neutrino mass is sufficiently small
compared to $E_0$  and assume that  there is no angular correlation between the electron and neutrino.
Here we imagine that the whole system is enclosed in a 
large box with a volume $V$. Hereafter we change the notation as $p_e \rightarrow p$
and $E_e \rightarrow E$.

In  Fermi's theory, the
four particles interact at a single point with a coupling constant $G_{F}$ so that
the matrix element is written as
\begin{eqnarray}
H_{fi}^{\prime} = G_{F} \int (\psi_{e}^{\dag}\mathcal{O}_L \psi_{\nu})( \psi_{p}^{\dag} \mathcal{O}_N \psi_{n}) dV, \label{matrix}
\end{eqnarray}
where $\psi_i$ is the wave function of each particle involved in the beta disintegration,
$\mathcal{O}_L$ and $\mathcal{O}_N$ are operators acting on 
the light particle's spin, nucleon's spin and isospin~(see, e.g., \citealt{feynman1958}
for a discussions on beta interaction). 
The wave function of the light particles can be evaluated at $r\sim 0$ because
their de Broglie wavelengths are much larger than the nuclear size.
When the light particles do not carry off orbital angular momentum with respect to 
the central nucleus,  the wave function of each light particle at $r=0$
is just a normalization factor  of $V^{-1/2}$ with a Coulomb correction for 
the electron's wave function. 
Thus the square of the matrix element can be written as 
\begin{eqnarray}
|H_{fi}^{\prime}|^2 = \frac{G_{F}^2 }{V^2} F(Z,E) |\mathcal{M}_N|^2, \label{matrix2}
\end{eqnarray}
where $F(Z,E)$ is the Coulomb correction factor, $Z$ is the proton number of
the daughter nucleus, and 
$\mathcal{M}_{N}$ is the nuclear matrix element.
The transitions described here are  {\it allowed} transition.
More specifically, allowed transitions are transitions which satisfy  both conditions that
the light particles don't carry off orbital
angular momentum and the parity of the nucleus does not change via its disintegration.
Otherwise the transition  is a {\it forbidden} transition.\footnote{We employ Konopinski's 
classification of beta decay~\citep{konopinski1966}.}
Because the population of allowed transitions is larger and because of their simplicity,
we focus on allowed transitions in this and the next sections.  We will discuss the role of forbidden transitions in \S 4.

Integrating 
Eq.~(\ref{fermi}) over the accessible phase space, 
the mean-life of a beta-unstable nuclide with the disintegration 
energy of $E_0$ is obtained as
\begin{eqnarray}
\frac{1}{\tau} 
& = & \frac{ \left| \mathcal{M}_{N}\right|^{2}}{t_{F}} \int_{0}^{p(E_0)} dp F(Z,E)
p^{2} (E-E_{0})^{2} , \label{mean}
\end{eqnarray}
where  the variables in the integral are in units of $m_{e}$ and $c$  and
$t_{F} $ is the fundamental timescale of beta decay: 
\begin{equation}
t_{F}  \equiv  \frac{2\pi^{3}}{G_F^2} \frac{\hbar^7}{m_{e}^{5}c^{4}}
\approx  8610~{\rm s}.\nonumber
\end{equation}
Note that, although this fundamental timescale is a characteristic timescale of allowed beta decay, 
the lifetime of beta unstable nuclides spreads over many orders of magnitude because of 
the phase space factor of Eq.~(\ref{mean}).

The Coulomb correction factor in the matrix element is obtained by 
evaluating the electron's wave function at the nuclear radius $r_n$~\citep{Fermi1934}:
\begin{eqnarray}\label{CC}
F(Z,E) & \cong & \frac{|\psi_e(r_n)|^2_{Z}}{|\psi_e(r_n)|^2_{Z=0}},\\ \nonumber 
&= &\frac{2(1+s)}{[(2s!)^2]} (2p\rho)^{2s-2} e^{\pi\eta} \left|(s-1+i\eta)! \right|^2, 
\end{eqnarray}
where $\eta = Zq^2_e/\hbar v$, $v$ is the velocity of the electron, 
$\rho = r_{n}/(\hbar/m_{e}c)$, $s=(1-(Z\alpha)^{2})^{1/2}$,
$q_e$ is the electron charge, and
$\alpha\approx 1/137$ is the fine-structure constant.
For $E>1$, the Coulomb correction factor slowly increases 
with $E$ as $F(Z,E)\propto E^{2s-2}$. 

A simple form of the Coulomb correction factor is obtained in the non-relativistic limit of Eq.~(\ref{CC}), $\eta \gg1$ and $(Z\alpha)^2\rightarrow 0$:
\begin{eqnarray}
F_{N}(Z,E) = \frac{2\pi \eta}{1-\exp(-2\pi \eta)}.\label{CN}
\end{eqnarray} 
The Coulomb correction factor is unity
for $\eta \ll 1$ and approaches to $2\pi \eta$ for $\eta \gg 1$. 
This enhances the transition probability at lower energies.
At these energies the electron is pulled by the nucleus due to the Coulomb force
and the amplitude of the electron's wave function is larger near the nucleus.
As a result, the lifetime of beta unstable nuclei becomes shorter than the one estimated 
without the Coulomb correction and the dependence of the lifetime on $E_0$
is weakened.
Note that one can also obtain an identical form to Eq.~(\ref{CN})  by solving the Schr${\rm\ddot{o}}$dinger equation and
evaluating the electron's wave function at $r=0$ .

As the integral in Eq.~(\ref{mean}) is easily calculated 
for given $E_{0}$ and $Z$, comparative 
half-lives $ft_{1/2}$ are often used for comparison with the experimental data:
\begin{eqnarray}
 ft_{1/2} \equiv \frac{\ln 2}{\left| \mathcal{M}_{N}\right|^{2}} t_{F}.
 \end{eqnarray}
Although $ \mathcal{M}_{N}$ of each beta transition cannot be calculated
within Fermi's theory, $\mathcal{M}_N$ can be determined from
the measurements of the lifetime and the electron's spectrum. 
It is sufficient for our purpose to know the statistical distribution of this quantity.
For allowed transitions, the distribution of $ft_{1/2}$ 
is known to have a peak around $10^{5}\,$s
corresponding to $\left| \mathcal{M}_{N}\right|^2\sim 0.05$~(e.g. \citealt{blatt1958}), which
we take as a reference value in this paper.\footnote{For neutron and mirror nuclides
such as $^3$H, the comparative half-lives are $\sim 10^3$~s corresponding to
$|\mathcal{M}_{N}|^2\sim 1$. Such transitions are called as {\it superallowed} transitions.
These transitions are, however, absent in $r$-process material.}

One can show that $f$ attains simple forms in the following three regimes:
\begin{eqnarray} \label{f}
f(Z,E_{0}) =
 \left\{                                                                                                                                                                             
\begin{array}{ll} 
\frac{1}{30}E_{0}^{5}~~~~~~~~\,({\rm relativistic}:~E_{0} > 1 ),\\ 
\frac{16\sqrt{2}}{105} E_{0}^{7/2}~~~(\rm{non\,relativistic}:\\ 
~~~~~~~~~~~~~~~~~~~~~~~~~~~E_c < E_{0} < 1),\\
\frac{2\pi Z\alpha}{3}E_{0}^{3}~~~~~({\rm non\,relativistic~Coulomb}:\\
~~~~~~~~~~~~~~~~~~~~~~E_0 < {\rm min}(E_c,1)),\\ 
\end{array}                                                                                                                                                                        
\right.
\end{eqnarray}
where $E_c=(2\pi Z\alpha)^2/2$.
The non-relativistic regime exists only for $Z\lesssim 30$,
and thus, there is no such a regime in $r$-process material.
In previous a work
\cite{colgate1966ApJ} applied only the relativistic regime 
$\tau \propto E^{-5}$. However, as we will see later, the mean-lives of the  nuclei
are rather proportional to $E^{-4}$ or $E^{-3}$ on the relevant timescale of macronovae,
i.e., a few days.

\begin{figure}
\includegraphics[width=80mm]{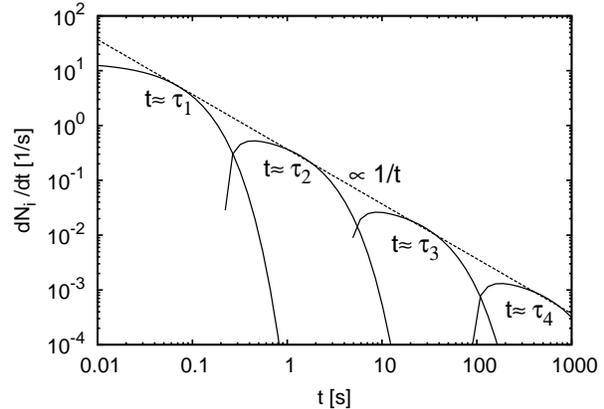}
\caption{A decay chain normalized by the total number of nuclei in the chain.
The dashed line depicts $e^{-1}/t$, where $e$ is Euler number.}
\label{fig:dN}
\end{figure}

In the context of macronovae, we are  interested in the relation between the lifetime
and the mean electron's energy since the
neutrinos don't  contribute to the  heat deposition in the merger ejecta. The fraction of energy of the electrons to the total energy 
is:
 \begin{eqnarray}
 \epsilon_{e} & \equiv & \frac{\langle E_{e} \rangle}{E_{0}}, \\ \label{eps}
 & = & \frac{1}{E_{0}} \int_{0}^{p_{0}}F(Z,p)p^2 E(E_{0}-E)^2 dp. \nonumber
 \end{eqnarray}
In the three regimes discussed earlier $\epsilon_{e}$  satisfies:
 \begin{eqnarray}\label{e}
 \epsilon_{e} =
 \left\{                                                                                                                                                                             
\begin{array}{ll}
1/2 &({\rm relativistic:}~E_{0} > 1 ),\\ 
1/3 &({\rm non\,relativistic:}~ (E_c < E_{0} <1),\\ 
1/4 &({\rm nr\,Coulomb:}~E_{0} < E_c),\\
\end{array}                                                                                                                                                                        
\right.
 \end{eqnarray}
 where we assumed $E_c<1$.

  \begin{figure*}
\includegraphics[width=80mm]{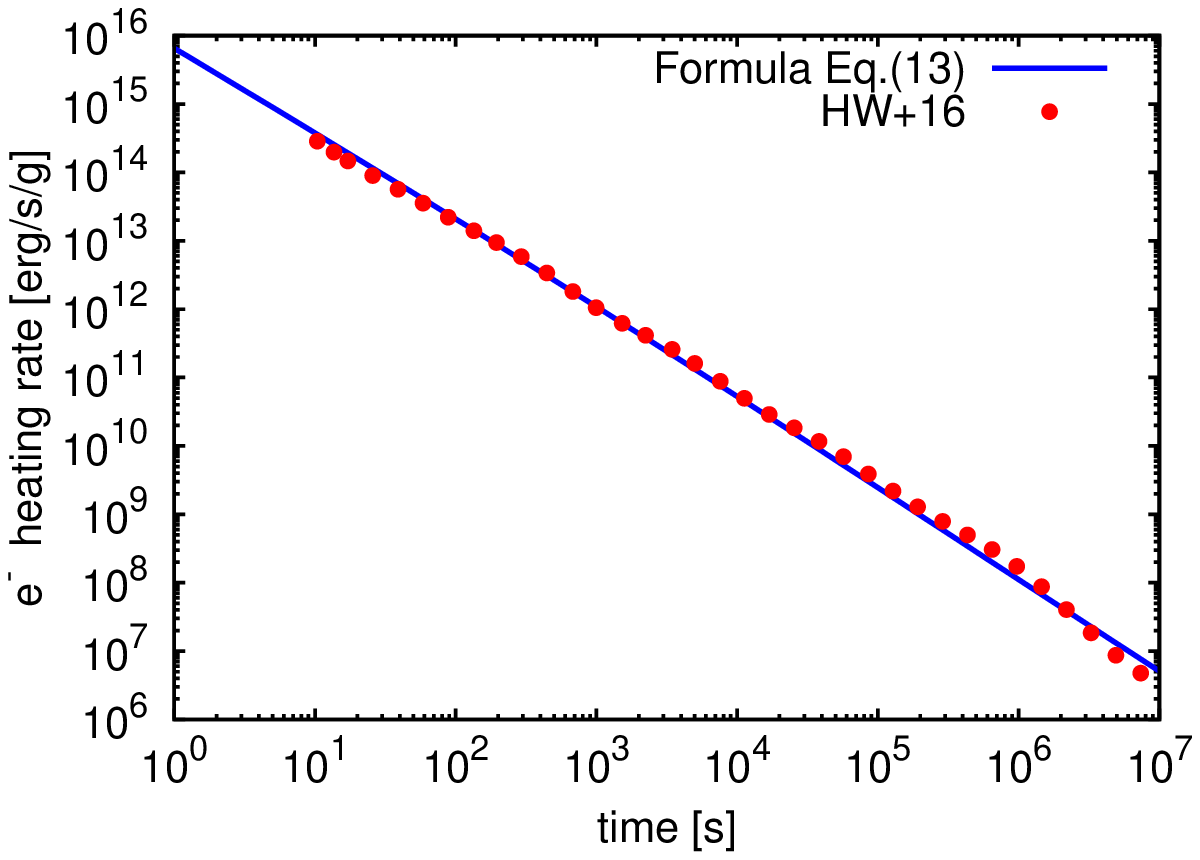}
\includegraphics[width=80mm]{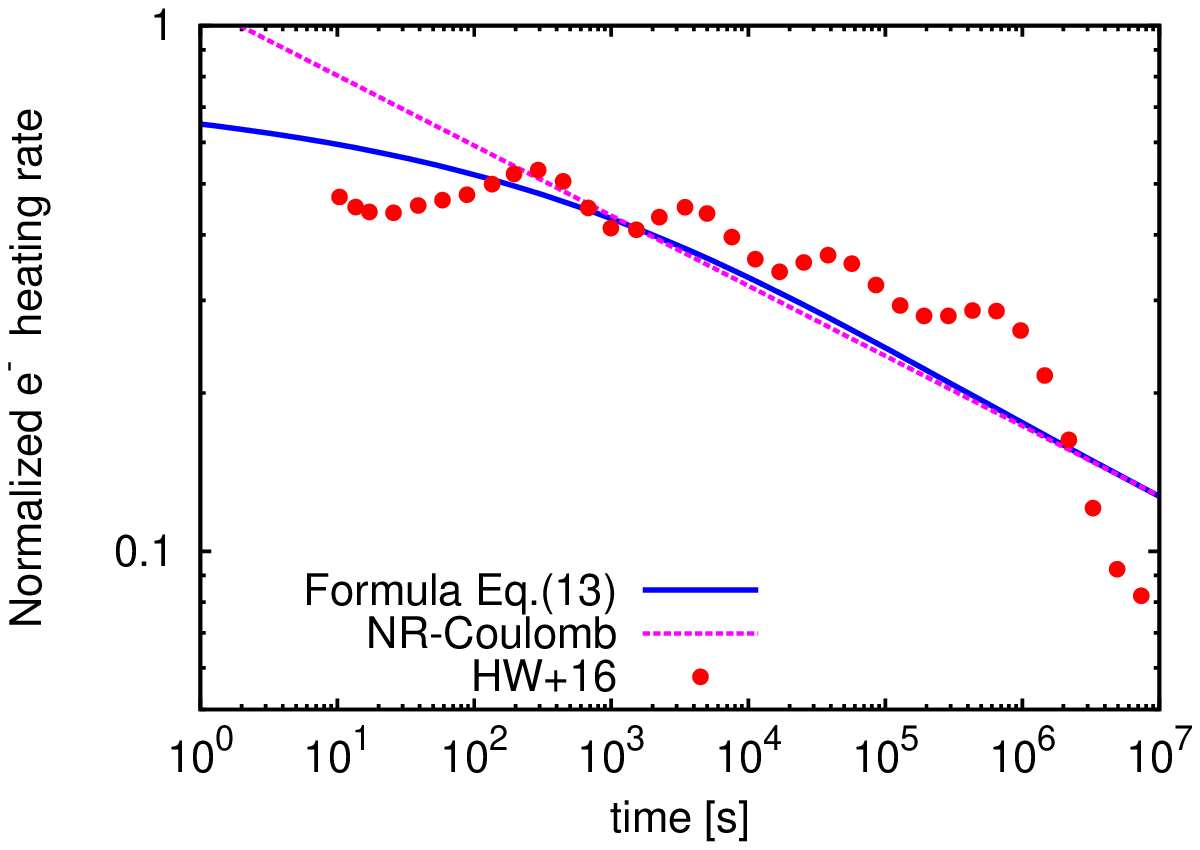}
\caption{The heating rate of the ideal chains of allowed beta decay by electrons. 
Left panel:  the specific heating rate derived by Eq.~(\ref{heat}).
Right panel:   the heating rate normalized by the relativistic regime 
of Eq.~(\ref{heat2}). Also shown in both panels is the electron heating rate 
taken from \citealt{hotokezaka2016MNRAS}, where
the heating rate is obtained by using Evaluated Nuclear Data File. 
Here we adopt $|\mathcal{M}_{N}|^2=0.05$,
$\langle A\rangle =200$, and $\langle Z\rangle =50$ for the analytic model.}
\label{fig:heat}
\end{figure*}

\section{The heating rate: the ideal-chains approximation} \label{sec:heat}
Neutron-rich nuclei produced via the $r$-process   
undergo beta decay towards the beta-stable valley without changing their mass number. 
A series of beta decays of nuclei in each mass number can be considered as a decay chain.
Here we consider ideal-chains of radioactive nuclei with a series of mean lives~($\tau_1 <\tau_2<\tau_3 <...$),
in which each chain conserves the total number of nuclei throughout the decay process and sufficiently many chains exist.
Within this approximation, the number of decaying nuclei in a logarithmic time interval is constant and 
the beta decays at a given time $t$ are dominated  
by nuclides with mean-lives of $\tau \sim t$~(see Fig.~\ref{fig:dN}).   This is, of course, valid for $t>\tau_{1}$,
where $\tau_{1}$ is the mean life of the first nuclide in a decay chain.
The heating rate per unit mass is then
\begin{eqnarray}
\dot{Q}(t) = -\sum_{i} \frac{E_{e,i}}{\langle A\rangle m_u}\frac{dN_i}{dt}
\approx \frac{e^{-1}}{\langle A \rangle m_{u}}\frac{\langle E_{e}(t) \rangle}{t},\label{heat}
\end{eqnarray}
where 
$\langle A \rangle$ is the mean mass number of the $r$-process material,
and $m_u$ is the atomic mass unit.  Note that $e$ is the Euler number,
which arises from the fact that the decay rate of each nuclide is
proportional to $e^{-t/\tau}$.
One can obtain $\dot{Q}$ by using Eq.~(\ref{mean}) and (11).

In the relativistic and non-relativistic Coulomb regimes, 
we can derive  simple explicit forms of Eq.~(\ref{heat}). 
As the lifetime of beta-unstable nuclides monotonically increases with decreasing $E_0$,
the relativistic regime is valid at early times and the non-relativistic Coulomb 
regime is valid  at late times.
More specifically, the relativistic  regime is   valid until
 $t_{\rm R}\approx 10^3\,{\rm s}\,(0.05/|\mathcal{M}_{N}|^{2})$
and the non-relativistic Coulomb regime is valid  after 
$t_{\rm NC}\approx 10^6\,{\rm s}\,(0.05/|\mathcal{M}_{N}|^{2})$.
Using Eqs.~(\ref{f}) and (\ref{e}), we obtain the heating rate in these regimes:
\begin{eqnarray}\label{q}
\dot{Q}(t) \approx 
 \left\{                                                                                                                                                                             
\begin{array}{ll}
1.2\cdot 10^{10}~{\rm \frac{erg}{s\cdot g}}~t_{\rm day}^{-\frac{6}{5}} \\ 
~~~~~~~~~\times \langle A \rangle^{-1}_{200}  \label{heat2}
\left(\frac{|\mathcal{M}_{N}|^{2}}{0.05}\right)^{-\frac{1}{5}}~~~~(t \lesssim t_{\rm R} ),\\ 
0.3\cdot 10^{10}~{\rm \frac{erg}{s\cdot g}}~t_{\rm day}^{-\frac{4}{3}} \\ 
~~~~\times \langle Z\rangle_{70}^{-\frac{1}{3}}\langle A \rangle_{200}^{-1}
\left(\frac{|\mathcal{M}_{N}|^{2}}{0.05}\right)^{-\frac{1}{3}}~(t  \gtrsim t_{\rm NC}),\label{heat2}\\ 
\end{array}                                                                                                                                                                        
\right. 
\end{eqnarray}
where $t_{\rm day}$ is time in units of a day, $\langle A\rangle_{200}$ is the mean mass number
normalized by $200$, and $\langle Z\rangle_{70}$ is the mean proton number normalized by $70$.
Note that the overall magnitude of the heating rate is determined by
the mean values of the nuclear quantities, $A$, $Z$, and $\mathcal{M}_{N}$.
These values should be constant within an order of magnitude, 
and thus, the magnitude of the heating rate does  not depend significantly on
the details of the abundance pattern of the $r$-process nuclei. 
Furthermore, we  emphasize that the formula of Eq.~(\ref{heat2}) is 
independent of the distribution of the nuclear decay energy.  

Figure \ref{fig:heat} depicts  the heating rate obtained from Eq.~(\ref{heat}) and
the one derived using a nuclear database~(\citealt{hotokezaka2016MNRAS}; see also similar
heating rates in \citealt{metzger2010MNRAS,goriely2011ApJ,roberts2011ApJ,
korobkin2012MNRAS, 
wanajo2014ApJ,lippuner2015ApJ}). 
We find that the heating rate based on the simple analytic formula reproduces  the one 
based on the database remarkably well.
In order to see more details, the right panel of Fig.~\ref{fig:heat}
shows the heating rates normalized to the values obtained for the relativistic regime  (Eq.~\ref{heat2}).
The normalized analytic heating rate (blue solid line) is flat  at early times and 
it approaches  the non-relativistic Coulomb regime (magenta dotted line) at late times.

It is worthy noting that the formula with the non-relativistic Coulomb limit 
reproduces the full heating rate after $10^3$\,s, 
even though it should be valid only  after $\sim 10^{6}$\,s. 
This can be understood as follows. The mean life is approximately proportional to 
$E_0^{-4}$ between the relativistic and the non-relativistic regimes,
and thus,  the energy generation rate evolves as $E_0/t \propto t^{-\frac{5}{4}}$. 
In addition, in this stage, $\epsilon_e$ changes from $1/2$ to $1/4$,
which approximately corresponds to $\epsilon_e \propto t^{-\frac{1}{9}}$. As a result,
the electron heating rate is $\propto t^{-1.35}$, which is quite similar to the one in the
non-relativistic Coulomb regime.

Note that  \cite{colgate1966ApJ} and \cite{metzger2010MNRAS}
assume 
that a nucleus that undergoes a radioactive decay reaches the valley of stability in a single step. In this case 
 the total number of radioactive nuclei decreases with time.
This  assumption is valid if the  radioactive nuclei are distributed just next to the
stable nuclei, i.e., at late times.  
Under such an assumption, the resulting heating rate declines more steeply
as $\propto t^{-1.4}$ in the relativistic regime. 
As we will discuss in the next section, 
the actual situation is in between these two assumptions.


\section{Deviation from our assumptions}\label{sec:discussion}
The analytic formula derived in the previous section reproduces  remarkably well the 
result based on the nuclear database. However, there are two 
important effects that have not been taken into account. Here we discuss
the role of these effects.
\subsection{The role of forbidden transitions}

\begin{figure}
\includegraphics[width=80mm]{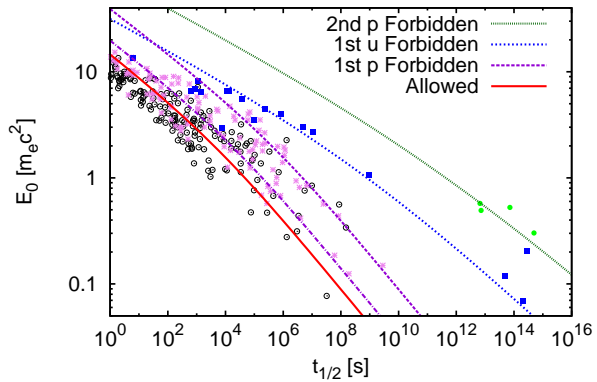}
\caption{Energy and half-life of beta unstable $r$-process nuclides.
Open circles, closes, filled squares, and filled circles are allowed,
first parity forbidden, first unique forbidden, and second parity forbidden transitions respectively.
Here the data points are taken from Evaluated Nuclear Data File ENDF/B-VII.1 
library~\citep{endf}.
Each curve depicts the theoretical expectation with a constant nuclear matrix element
of each type of transitions: the allowed~(red solid), 
the first parity forbidden~(magenta dashed and dot-dashed), 
the first unique forbidden~(blue dotted), and the second parity forbidden~(green).
Here we adopt $|\mathcal{M}_{N}|^2=0.05$
and $\langle Z\rangle =50$ for all analytic models
but $|\mathcal{M}_{N}|^2=0.01$ for the first unique transition. 
}
\label{fig:half}
\end{figure}

{\it Higher orbital-angular momentum transitions~(unique forbidden)}:
The light particles' wave function in the matrix element Eq.~(\ref{matrix}) 
can be expanded in a series of
spherical harmonics, of which
the $l$th term is proportional to $(Pr/\hbar)^{l}$, where 
$\vec{P}$ is the total momentum of the light particles.
The $l$th transition corresponds to the transition in which 
the light particles carry off orbital angular momentum of $l\hbar$. 
This expansion converges rapidly on the energy scale of beta decay on 
the length scale of nucleus $r_n \sim fm$. 
As a result, the $l$th transition probability  
is suppressed by a factor of $(Pr_{n}/\hbar)^{2l}\lesssim (0.1)^{2l}$.
For first unique forbidden transitions,  an additional shape factor
$2(p_{\nu}^2+p_e^2)$ should be multiplied in the electron spectrum of Eq.~(\ref{mean}).
This shape factor results in $\tau \propto E_0^{-5}$ in the non-relativistic Coulomb regime,
which can be seen in Fig.~\ref{fig:half} (a blue dotted line). Even though 
the number of beta unstable nuclides that disintegrate mainly via 
unique forbidden transitions is small, they may play a
role by increasing  the heating rate after a few hours.

{\it Relativistic transitions (parity forbidden)}:
Some interactions  mix 
the large and small components of Dirac spinor of the nucleon
in the matrix element Eq.~(\ref{matrix}).
A transition due to such an interaction  is a {\it parity forbidden} transition as it  
changes the nucleus' parity without removing the orbital
angular momentum (see magenta crosses in Fig.~\ref{fig:half}). 
The corresponding amplitude is suppressed 
by  a factor of $\mathcal{O}(v_{n}/c)$ or $\mathcal{O}(Z\alpha)$
compared to allowed transitions.
Here the velocity of nucleons $v_{n}$ is typically $v_n \sim 0.1c$.
As a result, the probability of these transitions is lower than the allowed ones
by a factor of $\mathcal{O}(v_{n}^2/c^2)$ or $\mathcal{O}(Z^2\alpha^2)$.
The theoretical curves of the first order parity forbidden transitions are  
shown  as the dashed and the dot-dashed lines in Fig.~\ref{fig:half}.
Here we use a suppression factor of $(v_n/c)^2\approx 0.01$ and $(Z\alpha)^2\approx 0.25$,
respectively. The first order parity forbidden transitions have an
electron spectral shape that is similar to the allowed transitions. 
As one can see in this figure,
the curves of these transitions have the same shapes to the allowed one
with a constant shift in the half-life. The existence of these transitions in addition to
the allowed ones increases the heating rate. 
The lifetimes of second order parity forbidden 
transitions, in which angular momentum of $\hbar$ is carried off by
the light particles, are too long to be relevant for the macronova heating rates
(see green points in Fig.~\ref{fig:half}).

\subsection{Deviation from the ideal-chains approximation}
\begin{figure}
\includegraphics[width=80mm]{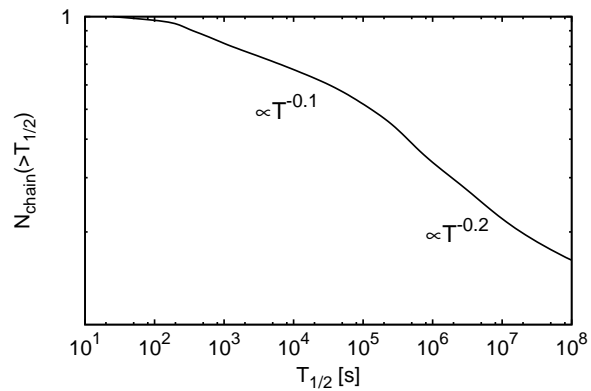}
\caption{The cumulative distribution of the lifetime of decay chains.
$T_{1/2}$ is the sum of the half-lives of beta unstable 
nuclides of a decay chain.}
\label{fig:chain}
\end{figure}

Beta decay chains terminate when they reach stable nuclides. Once this happens
these terminated chains don't contribute to the heating rate any more.
The overall  lifetime of a chain, $T_{1/2}$, can be estimated from the sum of the half-lives of 
nuclides in the chain.
The cumulative distribution of the chains for $A=90$--$210$ 
as a function of the chains' lifetime is shown in Fig.~\ref{fig:chain}.
The number of the chains begins to decrease slowly as $\propto T_{1/2}^{-0.1}$ at $\sim 100$\,s.
After about $10$ days it decreases slightly faster as $\propto T_{1/2}^{-0.2}$.
This steep decline at late times due to the termination of the decay chains
 is consistent with the assumption made by 
\cite{colgate1966ApJ} and \cite{metzger2010MNRAS}.

In summary, the contribution of forbidden transitions to the heating rate 
slightly increases the  heat generation at late times. On the contrary, at the same time,
the termination of the beta decay chains slightly decreases it. 
As a result, the combined effects on the heating rate somehow cancel out.
Note that these corrections to the heating rate depend on the actual abundance 
distribution of the chains.

\section{Conclusion and Discussion}\label{sec:conclusion}
We derive an analytic form of the macronova heating rate 
by considering statistical assembly of radioactive $r$-process
nuclides and Fermi's theory of beta decay. 
The resulting analytic formula reproduces the heating rate derived from
the nuclear database remarkably well. 
Within the assumption
that the ideal decay chains of allowed beta transitions generate
radioactive heats, we show that the heating rate evolves
as $\propto t^{-6/5}$ at early times  and $\propto t^{-4/3}$
at late times. The overall magnitude of heating rate is determined by
the mean value of the nuclear matrix elements, mass and atomic 
number of beta unstable nuclides involved in the decay chains.

We discuss the role of forbidden transitions and the deviation from
the ideal-chains approximation. The former slightly increases the
heating rate at late times and the latter slightly decreases it.
As a result, these corrections somehow cancel out with each other.

The robust and simple form of the heating rate suggests that 
observations of the  late-time bolometric macronova light curve
can provide  and observational evidence that is is driven by a radioactive 
decay of  $r$-process material. Furthermore, determination of the bolometric luminosity will enable us to 
estimate the total amount of $r$-process nuclei produced in a merger.
Using the non-relativistic Coulomb regime of Eq.~(\ref{heat2}), the late-time bolometric light curve is written as
\begin{equation} 
L(t)\approx 10^{40}\,{\rm erg/s}\,t_{5{\rm day}}^{-\frac{4}{3}}M_{-2} \  {\rm erg/s} \ .
\end{equation}
where  $M_{-2}=0.01M_{\odot}$ is the ejecta mass. This expression is 
valid after the peak time given by
\begin{equation}
t_{\rm peak}\approx 5\, \,\kappa_{10}^{\frac{1}{2}}M_{-2}^{\frac{1}{2}} v_{0.2}^{-\frac{1}{2}}  {\rm days} \ ,
\end{equation} 
where  $v_{0.2}=0.2c$ is the ejecta velocity and $\kappa_{10}=10~{\rm cm^{2}/g}$ is 
the bound-bound opacity of $r$-process elements~\citep{kasen2013ApJ,tanaka2013ApJ}.

Confirming this behavior by observation  may be difficult because the light curve may have large fluctuation
due to the temperature and density dependent opacity~\citep{barnes2013ApJ,tanaka2013ApJ}.
However, as suggested in the context of supernovae~(\citealt {katz2013,nakar2016ApJ}), 
the time-weighted integral of the bolometric luminosity  after the peak provides a
more robust  estimate  the radioactive power in the ejecta. In this method,
the time-weighted integral of the bolometric luminosity should behave as
$\propto M\cdot t^{2/3}$. 

The bolometric luminosity that we derived  here is the total radioactive power
 emitted in the electrons. At late times, this power is not necessarily  thermalized in the ejecta
 ~(see \citealt{barnes2016} for a detailed study).
 The inefficiency of the electron thermalization may reduce the bolometric luminosity
  by a factor of $2$ on the macronova timescale. At the same time, we have ignored,  
 additional heating due to $\gamma$-rays, $\alpha$-particles
and fission fragments. The role of these decay products in the macronova heating 
is still under debate. For instance, it has been suggested that the heat generation by 
spontaneous fission and $\alpha$-decay can be comparable to or even larger than
the beta decay heating (see \citealt{hotokezaka2016MNRAS,barnes2016}).
We explore the role of these effects in the estimating the total amount of $r$-process material ejected in a macronva
from the integrated bolometric light curve
in a separate work. 

\vspace{0.5cm}
\section*{Acknowledgments}
We thank Michael Paul, Kohsaku Tobioka, and Shinya Wanajo for useful discussions. 
This research was supported by an ERC advanced grant (TReX) and by the  I-CORE 
Program of the Planning and Budgeting Committee and The Israel Science
Foundation (grant No 1829/12), and an ISF grant. 

\bibliographystyle{mn2e}

\end{document}